\documentclass[a4paper,10pt]{article}
\usepackage[utf8]{inputenc}
\def\dis{\displaystyle}
\def\del{\partial}
\def\noi{\noindent}
\def\d{\rm d}

\title{Virtual Black Holes in a  Third Quantized Formalism }
\author{ Yoshiaki Ohkuwa$^a$ \footnote{  ohkuwa@med.miyazaki-u.ac.jp (Y. Ohkuwa),
                    mirfaizalmir@googlemail.com (M. Faizal),
                    ytar@hi2.enjoy.ne.jp (Y. Ezawa).}
, Mir Faizal$^{b, c}$,  Yasuo Ezawa$^d$, \\ \\
$^a$Division of Mathematical Science, \\ Department of Social Medicine,\\ Faculty
of Medicine,
University of Miyazaki, \\ Kihara 5200,  Kiyotake-cho, Miyazaki, 
889-1692, Japan\\  \\ 
$^b$Department of Physics and Astronomy,\\
University of Lethbridge, \\
Lethbridge, Alberta, T1K 3M4, Canada\\\\ $^c$Irving K. Barber School of Arts and Sciences,
\\ University of British
Columbia - Okanagan, \\
  Kelowna,  British Columbia V1V 1V7, Canada\\\\
$^d$Department of Physics, Ehime university,\\ 2-5 Bunkyo-cho, Matsuyama, 
790-8577,
Japan \\ \\ 
}
\date{}
\begin{document}

\maketitle

\begin{abstract}
In this paper, we will analyse virtual black holes using the third quantization formalism. 
As the virtual black hole model depends critically on the assumption that the 
quantum fluctuations dominate the geometry  
of spacetime at Planck scale, we will analyse the quantum fluctuations 
for a black hole using third quantization. 
We will demonstrate that these quantum fluctuations 
  depend   on the factor ordering chosen. So, we will show that only certain values of 
the factor ordering parameter are consistent with  virtual black holes model of spacetime foam. 
\\

\noi
\\

\noi
\end {abstract} 

\section{Introduction}

  A universal prediction from almost all approaches to quantum gravity is that 
 the geometry of spacetime will 
be dominated by quantum fluctuations near Planck scale, such that it would not
be possible to analyse the geometric structures below Planck scale   
\cite{su}-\cite{su12}. Furthermore, as there is strong evidence of the existence 
of macroscopic black holes,  the virtual 
black holes are expected to form due to these quantum fluctuations of spacetime. 
This is because all  approaches to quantum gravity should produce  
classical general relativity, in the classical limit. Thus, the formation of black holes 
should be allowed in all approaches to quantum gravity, and hence, quantum fluctuations 
at Planck scale should produce virtual black holes, in all these approaches. 
So, even though we will study the virtual black holes using a specific approach, 
it is expected that such virtual black holes will also occur  using other approaches 
to quantum gravity. 
We will analyse the virtual black holes using the   Wheeler-DeWitt equation, as the 
 quantum theory of black holes has been studied  using 
the Wheeler-DeWitt equation \cite{Majumder}-\cite{q}. 
The mass of the black holes changes dynamically with time, so a variable relating to time is 
obtained from the mass of the black hole, and this Wheeler-DeWitt equation is constructed
using such a  variable. Now as the black holes are valid quantum states in this theory, 
they can also be produced from quantum fluctuations. Such quantum fluctuations in the geometry, 
which change the topology of spacetime would occur near Planck scale. 
So, it is expected that the spacetime at Planck scale would be filled by a sea of virtual 
black holes  \cite{hb}-\cite{bh}. 

To analyse the formation of virtual black holes, we would 
require a formalism in which the geometries would be dynamically produced,
and which would allow for topology changing processes to occur. 
Thus, we will use the third quantized formalism 
to analyse virtual black holes \cite{bh}. 
       This is because 
 second quantized  Wheeler-DeWitt equation is the functional $\rm Schr\ddot{o}dinger$ 
 equation  describing the quantum state of geometries,  
  \cite{DeWitt67}-\cite{Wheeler57}, and  
it is not possible to analyse the dynamic creation or annihilation of geometries, 
 or even a multi-geometry state, using this second quantized Wheeler-DeWitt equation.
 In fact, it is well known that it is not possible to analyse
 topology changing processes using the second quantized Wheeler-DeWitt equation.
 However, the creation and annihilation of virtual black holes is 
   a process that changes the topology of spacetime. Hence, it cannot be properly 
   analysed using a second quantized Wheeler-DeWitt equation.
 This is similar to the situation in first quantized formalism with 
 particle creation and annihilation, as it 
 is not possible to describe the dynamic creation or annihilation of particles, 
or even a free multi-particle state using 
 a first quantized single particle  $\rm Schr\ddot{o}dinger$  equation. 
 However, just as we can describe the dynamic creation  and annihilation of 
 particles in second 
 quantized formalism,  we can analyse the creation and annihilation of geometries 
 in the third quantized formalism  \cite{th}-\cite{th1}. 
 So, in the third quantized formalism, the Wheeler-DeWitt equation is not viewed 
 as a quantum mechanical functional 
 $\rm Schr\ddot{o}dinger$ equation describing the quantum evolution of the 
 wave function of geometries, but rather as a classical field equation. 
 This equation is third quantized, and the creation and annihilation operators
 thus obtained make it possible   
to create and annihilate different geometries. 
 Thus, the third quantized Wheeler-DeWitt equation is usually used for analysing the 
 multiverse
 \cite{mult}-\cite{mult1}, but here we will use it for analyzing 
 virtual black holes \cite{bh}.

  It may be noted that this third quantized model of virtual black holes has been
  used to motivate an interesting 
  solution to the problem of time in quantum gravity \cite{bh}. According to this proposal, the entropy of the universe would increase due to 
  the interaction of virtual black holes with all other particles. This is because the particles  would fall into these virtual black holes
  and be radiated as different particles. This would cause information to be lost, and this lost information would increase the 
  entropy of the universe. Thus, it would be possible to define a direction of time by identifying time with the increase of the entropy of the 
  universe.  The loss of quantum coherence through scattering off virtual black holes has also been studied \cite{hb}. 
  In this analysis, a quantum field on the $C$ metric, was analysed. As the $C$ metric has the same topology as a pair of 
  virtual black holes, it was argued that such processes can be used for understanding some important features of virtual black holes. 
  A scalar field theory was analysed on this metric, and it was explicitly demonstrated that there is a 
    loss of quantum coherence. This calculation was generalized to include higher spin field \cite{h1}. The  
    transmission coefficient were calculated and used for analysing the 
  the loss of quantum coherence of an incident field through scattering off virtual black holes.
  Virtual black holes in two dimensional quantum gravity have also been analysed \cite{h2}-\cite{2h}.  
  Virtual black holes have also been studied in the context using   generalized dilaton theories \cite{d1}. 
  The phenomenological aspects of virtual black holes have also been studied \cite{d2}.

  The  virtual black holes have also been used for explaining the end state of evaporation of real black holes \cite{ha}. 
  According to this model, the real black holes reduce in size due to Hawking radiation. This process continues till 
  they reach Planck size. At this point they are lost in a sea of virtual black holes. However, this model implicitly 
  relies on the assumption that the solutions to the Wheeler-DeWitt equation for a black hole is dominated 
  by quantum fluctuations when the black hole becomes sufficiently small. So, it is important to analyse the effects of quantum fluctuations on 
  the solutions of the Wheeler-DeWitt equation.   
 The fluctuations in the geometry of the universe has been studied using third quantization   \cite{ab}.
  The third quantized formalism has also been used for analysing fluctuations for  
   Brans-Dicke theories  \cite{ai}, 
 $f(R)$ gravity theories \cite{f}-\cite{f1}, and Kaluza-Klein theories  \cite{ia}. Third quantization has also been used for analysing 
  the string perturbative vacuum   \cite{st}. However, the fluctuations in the geometry of a black hole have never been analysed using the third 
  quantized formalism. It is important to analyse such fluctuations in the geometry of a black hole, as the virtual black hole model 
  of spacetime foam relies on the assumption that such fluctuations dominate the geometry at Planck scale. So, in this paper, we will analyse 
 the effect of such fluctuations. We will observe that such fluctuations depend critically on the factor ordering chosen. 
 Hence, only certain factor orderings 
 are consistent with virtual black hole model of spacetime foam.  It has been observed that factor ordering can have real physical 
 consequences  
  \cite{oo}-\cite{Ohkuwa-Faizal-Ezawa1}. 
So,  such a dependence of the physics of virtual black holes on the factor 
ordering chosen was something that was expected.  
We will demonstrate this to be the case explicitly in this paper.

In section 2, we will consider the third quantization of the black hole. 
In section 3, we will analyse the uncertainty relation for the black hole. 
In section 4, we will study the operator ordering for the Wheeler-DeWitt equation 
of the black hole. 
Conclusion will be given in section 5.

\section{Third Quantization of Black Hole}\label{1} 
In this section, we will analyse the third quantization of virtual black holes. 
This will be done by first analyzing the Wheeler-DeWitt equation for a real black hole, 
in the minisuperspace approximation. As the mass of the black hole changes
with time, such a mass can be 
used to obtain a suitable time variable, for such a system. Then the Wheeler-DeWitt equation 
can be constructed using this time variable. We will third quantize this 
second quantized Wheeler-DeWitt equation, and this
will allow the dynamic creation and annihilation 
of such black holes. 
Now as the theory will allow the   creation and annihilation of real 
black holes, it would be expected that the   creation and annihilation of virtual 
black holes will also occur at Planck scale, due to quantum fluctuations. 
Furthermore, as the time variable will be expressed in terms of the mass of the black hole, 
such quantum fluctuations can be expressed in terms of the mass of the black hole. 
So, we can 
  start with the classical Hamiltonian for a  
Schwarzschild black hole,   
$
H = \frac{p_a^2}{2a} + \frac{a}{2}  \ 
$
 \cite{Majumder}-\cite{q}, 
where   $a$ roughly corresponds to the size (mass) of the black hole and
$p_a$ is the momentum   conjugate to $a$.  Both $a$ and $p_a$ are related to  
$m$ and $p_m$ through a canonical transformation. 
Here, we have 
$m(t):=M(t,r)$ and $p_m(t):= \int_{-\infty}^\infty dr \, p_M(t,r)$, and so  
$m$ is related to the 
  the mass of the black hole, which is denoted by $M$. 
This relation holds for  a spherically 
symmetric spacetime, and so the physics of the system is expressed by  the mass
of the black hole. 
As we are interested in the process that the mass evaporates due to the Hawking radiation, 
it can be used to construct a suitable time variable for the black hole. 
The variable $a$ satisfies $ 0 < a \leq 2M $. 
We consider the course that the  black hole starts from the limit $a \rightarrow 2M$, 
becomes smaller owing to the Hawking radiation, and ends at the limit $a \rightarrow 0$. 
So let us define $b \equiv 2M-a$ , and 
 we regard $b$ as the time variable for the black hole.  
Now we can write the  
 Wheeler-DeWitt equation for the Schwarzschild black hole 
 as  \cite{Majumder}-\cite{q}, 
$$
\left[ {1 \over a^{p_o +1}} {{\d} \over {\d} a}
a^{p_o} {{\d} \over {\d} a} 
- (a-2M) \right] \psi (a)  = 0 \ ,             
$$
which is equivalent to 
$$
\left[ {{\d}^2 \over {\d} a^2}
+ {p_o \over a} {{\d} \over {\d} a} 
- (a^2-2M a) \right] \psi (a)  = 0 \ ,                   \eqno(2.1)
$$                                                
where  $p_o$ is the operator ordering parameter. 
Now we rewrite these equations by using our time variable $b$ as
$$
\begin{array}{ll}
&\biggl[ \dis{{1 \over (2M-b)^{p_o +1}} {{\d} \over {\d} (2M-b)}
(2M-b)^{p_o} {{\d} \over {\d} (2M-b)}} \\
&\qquad\qquad\qquad\qquad\qquad\qquad\qquad\quad - [(2M-b)-2M] \biggr] \psi (b)  = 0 \ , 
\end{array}            
$$
which is equivalent to 
$$
\left[ {{\d}^2 \over {\d} b^2}
- {p_o \over 2M-b} {{\d} \over {\d} b} 
- [(2M-b)^2-2M (2M-b)] \right] \psi (b)  = 0 \ .          \eqno(2.2)
$$ 
 The Lagrangian for the third quantization whose variation gives Eq. (2.2) is 
$$
\begin{array}{ll}
{\cal L}_{3Q} &= {1 \over 2}
\biggl[ \dis{(2M-b)^{p_o} \left( {{\d}\psi(b) \over {\d} b}\right)^2 } \\
&\qquad + (2M-b)^{p_o} [(2M-b)^2-2M (2M-b)] \psi (b)^2
\biggr] \ . 
\end{array}                                                 \eqno(2.3)
$$ 
Thus, by defining 
$$
S_{3Q} = \int {\d}b \ {\cal L}_{3Q} \ ,                      \eqno(2.4)
$$
we can get Eq. (2.2) from 
$ \delta S_{3Q} = 0 $ . 
The momentum canonically conjugate to $\psi (b)$ is defined as 
$$
\pi (b) = {\del {\cal L}_{3Q} \over \del \left(
{{\d} \psi (b) \over {\d} b} \right)} 
= (2M-b)^{p_o} {{\d} \psi (b) \over {\d} b} \ .               \eqno(2.5)
$$
The Hamiltonian for the third quantization can be written as  
$$
\begin{array}{ll}
{\cal H}_{3Q} &= \dis{\pi (b) {{\d} \psi (b) \over {\d} b}
-{\cal L}_{3Q}} \ , \\[5mm]
&=\dis{{1 \over 2}\biggl[{1 \over (2M-b)^{p_o}} \pi (b)^2} \\
&\qquad - (2M-b)^{p_o} [(2M-b)^2-2M (2M-b)] \psi (b)^2
\biggr] \ . 
\end{array}                                                  \eqno(2.6)
$$

Now we can   third quantize this theory  by imposing 
the following equal time commutation relation, 
$$
[{\hat \psi} (b) , {\hat \pi} (b) ] = i \ ,                  \eqno(2.7)
$$
where a hat denotes  that we are dealing with an   operator. 
If we take the $\rm Schr\ddot{o}dinger$ picture, we have 
the time-independent $c$-number $\psi$ for the operator 
${\hat \psi} (b)$ .  
Therefore, we can rewrite the operators as   
$$
{\hat \psi} (b) \rightarrow \psi \ , \qquad
{\hat \pi} (b) \rightarrow 
-i{\del \over \del \psi} \ .                                 \eqno(2.8)                             
$$ 
Thus, we   have the $\rm Schr\ddot{o}dinger$ equation for virtual black holes, 
$$
\begin{array}{ll}
&\dis{i{\del \Psi (b, \psi) \over \del b}} = {\hat {\cal H}}_{3Q} 
\Psi (b, \psi) \ , \\[5mm]
&\qquad\ \ {\hat {\cal H}}_{3Q}= 
\dis{{1 \over 2}\Biggl[- {1 \over (2M-b)^{p_o}} 
{\del^2 \over \del \psi^2}} \\  
&\qquad\qquad\qquad\quad - (2M-b)^{p_o} [(2M-b)^2-2M (2M-b)] \psi^2
\Biggr] \ . 
\end{array}                                                  \eqno(2.9)
$$
Here $\Psi (b, \psi )$ is the third quantized wave function of the black hole.

\section{Uncertainty Relation}\label{2}
In the previous section, we described the black holes using a third quantized formalism. 
In this section, we will analyse the uncertainty relation for such a model. 
It is important to analyse the uncertainty for black holes, as we expect the quantum fluctuations to dominate 
near the Planck scale for the virtual black hole model to be consistent. 
We assume that the solution to Eq. (2.9) has the Gaussian form 
\cite{ai}-\cite{f1}, \cite{Ohkuwa-Faizal-Ezawa1} 
$$
\Psi (b, \psi) = C {\rm exp} \left\{ -{1 \over 2}A(b)
[\psi-\eta (b)]^2 +i B(b)[\psi-\eta (b)]
\right\} \ ,                                                 \eqno(3.1)
$$
where $C$ is a real constant, and $A(b)=D(b)+iI(b)$. Here $A(b), B(b), \eta (b)$ 
must be determined from Eq. (2.9). 
Note that in order for Eq. (3.1) to satisfy Eq. (2.9) $D(b)$ and $I(b)$ are 
real functions, but $B(b)$ and  $\eta(b)$ are complex functions in general. 
We can define the inner product of two third quantized wave functions 
$\Psi_1$ and $\Psi_2$  as 
$$
\langle \Psi_1 , \Psi_2 \rangle 
 \equiv \int_{-\infty}^{\infty} \! d \psi \, \Psi_1^*(b,\psi)
 \Psi_2(b,\psi)          .                                   \eqno(3.2)
$$
Now we define  the inner product of the Gaussian form wave function as 
$$
\langle \Psi , \Psi \rangle 
=C^2 \sqrt{{\pi \over D(b)}} \exp E(b) \ ,                   \eqno(3.3)
$$
where $E(b)$ is a real function, such that
$$
\begin{array}{ll}
E(b)=&\dis{{1 \over 2[A(b)+A^* (b)]}} \\[5mm]
&\times \{ -A(b)A^* (b) [\eta (b)-\eta^* (b) ]^2 - [B(b)-B^* (b)]^2 \\[3mm]
&\quad -2i [A^* (b) B(b) + A(b) B^* (b)][\eta (b) - \eta^* (b)] \} \ . 
\end{array}
$$

Now we calculate the Heisenberg's uncertainty relation, and this is done by 
 defining the dispersion of $\psi$  as
$$
(\Delta \psi)^2 \equiv \langle \psi^2 \rangle
-\langle \psi \rangle^2 \ , \qquad
\langle \psi^2  \rangle 
\equiv {\langle \Psi ,  \psi^2 \Psi \rangle \over 
\langle \Psi , \Psi \rangle } \ .                            \eqno(3.4)
$$
From Eqs. (3.1), (3.2), (3.3) and (3.4),  we obtain  
$$
\langle \psi^2  \rangle = {1 \over 2D(b)} 
+ F^2 (b) \ , 
\quad \langle \psi \rangle = F(b) \ , \quad {\rm and} \quad
(\Delta \psi)^2 = {1 \over 2D(b)} \ ,                        \eqno(3.5)
$$
where $F(b)$ is a real function, such that 
$$
F(b)={A(b) \eta (b) +A^* (b) \eta^* (b)
+i [B(b)-B^* (b)] \over A(b) + A^* (b)} \ . 
$$
We define the dispersion of $\pi$  as
$$
(\Delta \pi)^2 \equiv \langle \pi^2 \rangle
-\langle \pi \rangle^2 \ , \qquad
\langle \pi^2  \rangle 
\equiv {\langle \Psi ,  \pi^2 \Psi \rangle \over 
\langle \Psi , \Psi \rangle } \ .                            \eqno(3.6)
$$
Then we can write 
$$
\begin{array}{ll}
&\dis{\langle \pi^2 \rangle 
= {D(b) \over 2}+{I^2 (b) \over 2D(b)}
+G^2 (b) , 
\quad \langle \pi \rangle = G(b)} \ ,  \\[5mm]     
{\rm and} \qquad \qquad \qquad \quad 
&\dis{(\Delta \pi)^2 
= {D(b) \over 2}+{I^2 (b) \over 2D(b)}} \ , 
\end{array}                                                  \eqno(3.7)
$$
where $G(b)$ is a real function, such that  
$$
G(b)={1 \over A(b) +A^* (b)} 
\{ A(b) B^* (b)+A^* (b) B(b)
-i A(b) A^* (b) [\eta (b)-\eta^* (b)] \} \ . 
$$
So,  we can write the Heisenberg's uncertainty  as
$$
(\Delta \psi)^2 (\Delta \pi)^2
={1 \over 4} \Biggl( 1+ {I^2 (b) \over D^2 (b)} 
\Biggr) \  .                                                 \eqno(3.8)
$$

If we substitute the assumption (3.1) to Eq. (2.9), 
we can obtain the equation for $A(b)$ 
as 
$$
-i {{\d} A(b) \over {\d} b}
=-{1 \over  (2M-b)^{p_o}} A(b)^2 
- (2M-b)^{p_o} [(2M-b)^2-2M (2M-b)] \ .                      \eqno(3.9)
$$
(Note that three complex equations for $A(b), B(b), \eta (b)$ can be obtained by 
comparing the order of $\psi$ in Eq. (2.9).  
However, Eq. (3.9) is enough 
for the calculation of the Heisenberg's uncertainty relation.)
Let us define  
$$
\sigma \equiv (2M-b)^{1-p_o}  = a^{1-p_o}\ ,  \quad (p_o \neq 1) \ .     \eqno(3.10)
$$
then we have 
$$
i(1-p_o ) {{\d} A(\sigma) \over {\d} \sigma}
+ A(\sigma)^2 
+ \sigma^{2p_o +2 \over 1-p_o }
- 2M \sigma^{2p_o +1 \over 1-p_o}
=0 \ .                                                      \eqno(3.11)
$$ 
Defining a function $u(\sigma)$ by the equation, 
$$
A(\sigma) = i(1-p_o) 
{{\d} \ {\rm ln} \ u(\sigma) \over {\d} \sigma} \ ,         \eqno(3.12)
$$
we obtain 
$$
\begin{array}{ll}
&{\dis {{\d}^2 u(\sigma) \over {\d} \sigma^2}
-{1 \over (1-p_o)^2}\sigma^{2p_o +2 \over 1-p_o } 
 u(\sigma) 
+{2M \over (1-p_o)^2}\sigma^{2p_o +1 \over 1-p_o } 
 u(\sigma)
= 0} \ , \\
{\rm namely}\qquad& \\
&{\dis {{\d}^2 u(\sigma) \over {\d} \sigma^2}
-{1 \over (1-p_o)^2}\sigma^{2p_o +1 \over 1-p_o } (a-2M)
 u(\sigma) = 0} \ . 
\end{array}                                                    \eqno(3.13)
$$
These equations are too complicated to be solved analytically. 
However, since we are interested in the late time limit 
($b \rightarrow 2M , \ a \rightarrow 0$), 
we can neglect the second term of the first equation in Eqs. (3.13) in this limit.
So we obtain  
$$
{{\d}^2 u(\sigma) \over {\d} \sigma^2} 
+{2M \over (1-p_o)^2}\sigma^{2p_o +1 \over 1-p_o } 
 u(\sigma)
= 0 \ .                                                     \eqno(3.14)
$$
We can solve this equation  using a Bessel function as 
$$
u(\sigma) = \sigma^{1 \over 2} {\cal B}_{1-p_o \over 3}
\left(
{2 \over 3} \sqrt{2M}  \sigma^{3 \over 2(1-p_o)}
\right) \ ,                                                 \eqno(3.15)
$$
where ${\cal B}$ is a Bessel function that satisfies \cite{Abramowitz-Stegun}
$$
{{\d}^2 {\cal B}_{1-p_o \over 3}(z) \over {\d} z^2}
+{1 \over z}{{\d} {\cal B}_{1-p_o \over 3}(z) \over {\d} z}
+\left( 1-{\bigl( {1-p_o \over 3} \bigr)^2 \over z^2}
\right) 
{\cal B}_{1-p_o \over 3}(z) =0 \ .                          \eqno(3.16)
$$
So, the general solution to Eq. (3.14), can be written as 
$$
u(\sigma) = c_J \sigma^{1 \over 2} J_{1-p_o \over 3}
\left( {2 \over 3} \sqrt{2M}  
\sigma^{3 \over 2(1-p_o)} \right)
+ c_Y \sigma^{1 \over 2} Y_{1-p_o \over 3}
\left( {2 \over 3} \sqrt{2M}  
\sigma^{3 \over 2(1-p_o)}  \right) 
\ ,                                                        \eqno(3.17)
$$
where $c_J$ and $c_Y$ are arbitrary complex constants and 
$J_{1-p_o \over 3}$ and $Y_{1-p_o \over 3}$ are Bessel functions.  

Now let us define 
$$
z \equiv {2 \over 3} \sqrt{2M} \, \sigma^{3 \over 2(1-p_o)}
= {2 \over 3} \sqrt{2M} (2M-b)^{3 \over 2}
= {2 \over 3} \sqrt{2M} \, a^{3 \over 2} \ ,                  \eqno(3.18)
$$
then Eq. (3.17) becomes   
$$
u(z) = \biggl( {z \over {2 \over 3} \sqrt{2M} } 
\biggr)^{1-p_o \over 3} 
[ c_J J_{1-p_o \over 3} (z) + c_Y Y_{1-p_o \over 3} (z) ] 
\ .                                                        \eqno(3.19)
$$
We can obtain from Eqs. (3.12), (3.18), (3.19)  
$$
\begin{array}{ll}
A(z) &= \dis{ i(1-p_o) {{\d} z \over {\d} \sigma} 
{{\d}\ {\rm ln} \ u(z) \over {\d} z} }\\[5mm]
&= \dis{ i \, \sqrt{2M}
\left( {z \over {2 \over 3} \sqrt{2M}} \right)^{2p_o+1 \over 3}
{c_J J_{-2-p_o \over 3} (z) + c_Y Y_{-2-p_o \over 3} (z) \over 
c_J J_{1-p_o \over 3} (z) + c_Y Y_{1-p_o \over 3} (z)} }
\ . 
\end{array}                                                \eqno(3.20)
$$
Here we have used \cite{Abramowitz-Stegun} 
$$
{{\d} {\cal B}_{1-p_o \over 3} (z) \over {\d} z} 
= {\cal B}_{-2-p_o \over 3} (z) 
-{1-p_o \over 3z}{\cal B}_{1-p_o \over 3} (z) \ .          \eqno(3.21) 
$$

As $A(z) = D(z) + i I(z)$ , we obtain    
$$
D(z)=\dis{{-i \, \sqrt{2M}
\left( {z \over {2 \over 3} \sqrt{2M}} \right)^{2p_o +1 \over 3} 
\over 
\pi z \vert c_J J_{1-p_o \over 3} (z) 
+ c_Y Y_{1-p_o \over 3} (z) \vert^2}
(c_J c^*_Y - c^*_J c_Y)} \ ,                                \eqno(3.22)
$$
where we have used  \cite{Abramowitz-Stegun}
$$
J_{1-p_o \over 3} (z) Y_{-2-p_o \over 3} (z)
-J_{-2-p_o \over 3} (z) Y_{1-p_o \over 3} (z)
= {2 \over \pi z} \ .                                      \eqno(3.23)
$$
(Note that $c_J c^*_Y - c^*_J c_Y$ is a pure imaginary number.) 
We can obtain
$$
\begin{array}{ll}
I(z)=&\dis{{\sqrt{2M} \over 2} 
\left( {z \over {2 \over 3} \sqrt{2M}} \right)^{2p_o+1 \over 3}
\over   
\vert c_J J_{1-p_o \over 3} (z) 
+ c_Y Y_{1-p_o \over 3} (z) \vert^2} \\[6mm]
&\times 
\biggl[ 2\vert c_J \vert^2 J_{-2-p_o \over 3}(z) J_{1-p_o \over 3}(z)
+2\vert c_Y \vert^2 Y_{-2-p_o \over 3}(z)Y_{1-p_o \over 3}(z)  \\[3mm]
&\quad +(c_J c_Y^* + c_J^* c_Y)
\Bigl( J_{-2-p_o \over 3}(z) Y_{1-p_o \over 3}(z) \\[3mm]
&\qquad \qquad \qquad \quad \quad
+ J_{1-p_o \over 3}(z) Y_{-2-p_o \over 3}(z) \Bigr)
 \biggr] \ .                          
\end{array}                                                \eqno(3.24)
$$
So, assuming $c_J c^*_Y - c^*_J c_Y \neq 0$  
(note that this means both of $c_J ,  c_Y$ are nonzero), 
we have 
$$
\begin{array}{ll}
\dis{I(z)^2 \over D(z)^2}
=&-\dis{\pi^2 z^2 \over 4 (c_J c^*_Y - c^*_J c_Y)^2} \\[6mm]
&\times
\biggl[ 2\vert c_J \vert^2 J_{-2-p_o \over 3}(z) J_{1-p_o \over 3}(z)
+2\vert c_Y \vert^2 Y_{-2-p_o \over 3}(z)Y_{1-p_o \over 3}(z)  \\[3mm]
&\quad +(c_J c_Y^* + c_J^* c_Y)
\Bigl( J_{-2-p_o \over 3}(z) Y_{1-p_o \over 3}(z) \\[3mm]
&\qquad \qquad \qquad \quad \quad
+ J_{1-p_o \over 3}(z) Y_{-2-p_o \over 3}(z) \Bigr)
 \biggr]^2 \ . 
\end{array}                                                \eqno(3.25)
$$
If we substitute Eq. (3.25) to Eq. (3.8), 
we can have the Heisenberg's uncertainty relation. 
We can use this uncertainty relation to discuss the   
behavior of the black hole at the last stage of its evaporation.

At the early time limit ( $b \rightarrow 0 , \ a \rightarrow 2M, 
\ \sigma \rightarrow (2M)^{1-p_o}$ ), so we  
  change the variable from $\sigma$ to $\alpha$,  
$$
\sigma =(2M)^{1-p_o} +\alpha \ .                          \eqno(3.26)
$$
This limit means $\alpha \rightarrow 0$, and we should consider 
only $O(\alpha^1)$ . Then we obtain from  Eqs. (3.13) and Eq. (3.26), 
$$
\begin{array}{ll}
&{\dis {{\d}^2 u(\alpha) \over {\d} \alpha^2}
-{1 \over (1-p_o)^2}[(2M)^{1-p_o} + \alpha ]^{2p_o +2 \over 1-p_o } 
 u(\alpha)} \\
&{\dis \qquad\quad +{2M \over (1-p_o)^2}[(2M)^{1-p_o} + \alpha ]^{2p_o +1 \over 1-p_o } 
 u(\alpha)} \\
\approx &{\dis {{\d}^2 u(\alpha) \over {\d} \alpha^2}
-{1 \over (1-p_o)^2}(2M)^{2p_o+2}
\biggl( \left[ 1+{2p_o+2 \over 1-p_o}{\alpha \over (2M)^{1-p_o}} \right]} \\
&{\dis \qquad\qquad\qquad\qquad\qquad\qquad\qquad
-  \left[ 1+{2p_o+1 \over 1-p_o}{\alpha \over (2M)^{1-p_o}} \right] \biggr)
u(\alpha)} \\
=&{\dis {{\d}^2 u(\alpha) \over {\d} \alpha^2}
-{(2M)^{3p_o+1} \over (1-p_o)^3}\alpha u(\alpha)\ =\  0} \ . 
\end{array}                                                   \eqno(3.27)
$$
This last equation has the solution 
$$
u(\alpha) = \alpha^{1 \over 2} {\cal B}_{1 \over 3}
\left( {2i \over 3}{(2M)^{3p_o+1 \over 2} \over (1-p_o)^{3 \over 2}}
\alpha^{3\over 2} \right)
\qquad (p_o <1)  \ ,                                               \eqno(3.28)
$$
and 
$$
u(\alpha) = \alpha^{1 \over 2} {\cal B}_{1 \over 3}
\left( {2 \over 3}{(2M)^{3p_o+1 \over 2} \over (p_o-1)^{3 \over 2}}
\alpha^{3\over 2} \right)
\qquad (p_o >1)   \ ,                                              \eqno(3.29)
$$
where ${\cal B}$ is the Bessel function. 
 
When $p_o <1$ , if we write 
$$
z =  {2i \over 3}{(2M)^{3p_o+1 \over 2} \over (1-p_o)^{3 \over 2}} 
\alpha^{3\over 2}  \ ,                                             \eqno(3.30)
$$
the general solution to the last equation of Eqs. (3.27) is 
$$
u(z) = \left(
-{3i \over 2}{(1-p_o)^{3 \over 2} \over (2M)^{3p_o+1 \over 2}} z 
\right)^{1 \over 3}
[c_J J_{1 \over 3} (z) + c_Y Y_{1 \over 3} (z) ] \ .               \eqno(3.31)
$$
Therefore, we obtain from Eqs. (3.12) and (3.31) 
$$
A(z) = - \left( {-3i \over 2} \right)^{1 \over 3}
(2M)^{3p_o+1 \over 3} z^{1 \over 3}
{c_J J_{-{2 \over 3}} (z) + c_Y Y_{-{2 \over 3}} (z) \over 
c_J J_{1 \over 3} (z) + c_Y Y_{1 \over 3} (z)}      \ ,            \eqno(3.32)
$$
where we have used the similar equation as Eq. (3.21). 
Since $A(z)=D(z)+iI(z)$, after a short calculation we can obtain
$$
{I(z)^2 \over D(z)^2} =
-{(\vert c_J \vert^2 p_- +\vert c_Y \vert^2 q_- 
+c_J c_Y^* r_- + c_J^* c_Y s_- )^2 \over
(\vert c_J \vert^2 p_+ +\vert c_Y \vert^2 q_+ 
+c_J c_Y^* r_+ + c_J^* c_Y s_+ )^2}
\ ,                                                                \eqno(3.33)
$$
where
$$
\begin{array}{ll}
p_{\pm} &= J_{-{2 \over 3}}(z) J_{1 \over 3}(-z)
\pm J_{1 \over 3}(z) J_{-{2 \over 3}}(-z) \\
q_{\pm} &= Y_{-{2 \over 3}}(z) Y_{1 \over 3}(-z)
\pm Y_{1 \over 3}(z) Y_{-{2 \over 3}}(-z) \\
r_{\pm} &= J_{-{2 \over 3}}(z) Y_{1 \over 3}(-z)
\pm J_{1 \over 3}(z) Y_{-{2 \over 3}}(-z) \\
s_{\pm} &= J_{1 \over 3}(-z) Y_{-{2 \over 3}}(z)
\pm J_{-{2 \over 3}}(-z) Y_{1 \over 3}(z)  \ . 
\end{array}
$$
From these equations and Eq. (3.8), 
we will obtain the Heisenberg uncertainty relation at the early time limit 
when $p_o <1$ in the next section.

When $p_o >1$ , if we write 
$$
z =  {2 \over 3}{(2M)^{3p_o+1 \over 2} \over (p_o-1)^{3 \over 2}} 
\alpha^{3\over 2}  \ ,                                             \eqno(3.34)
$$
the general solution to the last equation of Eqs. (3.27) is 
$$
u(z) = \left(
{3 \over 2}{(p_o-1)^{3 \over 2} \over (2M)^{3p_o+1 \over 2}} z 
\right)^{1 \over 3}
[c_J J_{1 \over 3} (z) + c_Y Y_{1 \over 3} (z) ] \ .               \eqno(3.35)
$$
Therefore, we obtain from Eqs. (3.12) and (3.35) 
$$
A(z) = -i \left( {3 \over 2} \right)^{1 \over 3}
(2M)^{3p_o+1 \over 3} z^{1 \over 3}
{c_J J_{-{2 \over 3}} (z) + c_Y Y_{-{2 \over 3}} (z) \over 
c_J J_{1 \over 3} (z) + c_Y Y_{1 \over 3} (z)}      \ ,            \eqno(3.36)
$$
where we have used the similar equation as Eq. (3.21). 
After the similar calculation as in Eq. (3.25) we have 
$$
\begin{array}{ll}
\dis{I(z)^2 \over D(z)^2}
=&-\dis{\pi^2 z^2 \over 4 (c_J c^*_Y - c^*_J c_Y)^2} \\[6mm]
&\times
\biggl[ 2\vert c_J \vert^2 J_{-{2 \over 3}}(z) J_{1 \over 3}(z)
+2\vert c_Y \vert^2 Y_{-{2 \over 3}}(z)Y_{1 \over 3}(z)  \\[3mm]
&\quad +(c_J c_Y^* + c_J^* c_Y)
\Bigl( J_{-{2 \over 3}}(z) Y_{1 \over 3}(z)
+ J_{1 \over 3}(z) Y_{-{2 \over 3}}(z) \Bigr)
 \biggr]^2 \ , 
\end{array}                                                \eqno(3.37)
$$
which will be used to obtain the Heisenberg uncertainty 
relation at the early time limit 
when $p_o >1$ .

\section{Operator Ordering}\label{b}
In this section, we will analyse the effect of operator ordering on the quantum fluctuations for the black hole. 
At late times namely when 
$b \rightarrow 2M , \ a \rightarrow 0$ i.e.,  
$z \rightarrow 0$ from Eq. (3.18), we should  
divide the cases by the value of the operator ordering parameter 
$p_o$ . 
For example, when we choose  
$$
p_o = 2  , \                                              \eqno(4.1)
$$  
as we obtain at late times  \cite{Abramowitz-Stegun} 
$$
\begin{array}{ll}
\dis{ J_{-{4 \over 3}}(z)} 
&\sim \dis{- {1 \over 3\Gamma \Bigl( {2 \over 3} \Bigr)}
 \left( {z \over 2} \right)^{-{4 \over 3}} ,
\qquad J_{-{1 \over 3}}(z) 
\sim {1 \over \Gamma \Bigl( {2 \over 3} \Bigr)}
\left( {z \over 2} \right)^{-{1 \over 3}}
 },  \\[6mm]
\dis{Y_{-{4 \over 3}}(z)} 
&\sim \dis{ {1 \over 3\sqrt{3} \Gamma \biggl({2 \over 3} \biggr)}  
\left( {z \over 2} \right)^{-{4 \over 3}} , 
\qquad Y_{-{1 \over 3}}(z) 
\sim  -{1 \over  \sqrt{3} \Gamma \Bigl( {2 \over 3}\Bigr)}
\left( {z \over 2} \right)^{-{1 \over 3}}
} ,  
\end{array}                                                \eqno(4.2)                         
$$
we can obtain from Eq. (3.25) 
$$
\begin{array}{ll}
\dis{I(z)^2 \over D(z)^2} &\sim 
\dis{ -{\pi^2 \over 9 \left( \Gamma \Bigl( {2 \over 3}\Bigr) \right)^4
 (c_J c^*_Y - c^*_J c_Y)^2}
} \\[6mm]
&\qquad \dis{\times \biggl[ -2\vert c_J \vert^2 -{2 \over 3} \vert c_Y \vert^2
+{2 \over \sqrt{3}} (c_J c_Y^* + c_J^* c_Y)
\biggr]^2 
\left( {z \over 2} \right)^{-{4 \over 3}}
} \\[6mm]
&\rightarrow \infty \ . 
\end{array}                                                \eqno(4.3)
$$
This and  Eq. (3.8) show  that at the end stage of evaporation  of the black hole  i. e., in the limit 
 $a \rightarrow 0$,   the quantum  fluctuations dominate the black hole. 
This is what is required in the virtual black hole model. 
It may be noted that here $a$ is related to the mass of the black hole. 
 Thus, as the mass of the black hole becomes small, quantum fluctuations
increase, and in the limit  $a \rightarrow 0$, the black hole gets lost in a sea of virtual black holes produced by 
quantum fluctuations.

However, a different choice of operator ordering can produce a different result. 
For example,  when we choose 
$$
p_o = 0 , \                                               \eqno(4.4)
$$
as we obtain at late times  \cite{Abramowitz-Stegun}
$$
\begin{array}{ll}
\dis{ J_{-{2 \over 3}}(z)} 
&\sim \dis{ {1 \over \Gamma \Bigl( {1 \over 3} \Bigr)}
 \left( {z \over 2} \right)^{-{2 \over 3}} ,
\qquad J_{1 \over 3}(z) 
\sim {3 \over \Gamma \Bigl( {1 \over 3} \Bigr)}
\left( {z \over 2} \right)^{1 \over 3}
 },  \\[6mm]
\dis{Y_{-{2 \over 3}}(z)} 
&\sim \dis{ {1 \over \sqrt{3} \Gamma \Bigl({1 \over 3} \Bigr)}  
\left( {z \over 2} \right)^{-{2 \over 3}} , 
\qquad Y_{1 \over 3}(z) 
\sim  -{ \Gamma \Bigl( {1 \over 3}\Bigr) \over \pi}
\left( {z \over 2} \right)^{-{1 \over 3}}
} ,  
\end{array}                                                \eqno(4.5)                         
$$
we can have  
$$
\begin{array}{ll}
\dis{I(z)^2 \over D(z)^2} &\sim 
-\dis{{1 \over  (c_J c^*_Y - c^*_J c_Y)^2}
\biggl[ {2 \over \sqrt{3}} \vert c_Y \vert^2
+ (c_J c_Y^* + c_J^* c_Y)
\biggr]^2} \\[6mm]
&\sim O(1) \ . 
\end{array}                                                \eqno(4.6)
$$
This and  Eq.(3.8) indicate  that at the end stage of evaporation  of the black hole, 
i. e., in the limit $a \rightarrow 0$ , 
the black hole would become classical in the sense 
that the quantum fluctuations become minimum. Thus, at the end stage of the evaporation of the black hole, 
virtual black holes cannot be produced from quantum fluctuations, as the  quantum fluctuations become minimum
at this stage. 
Thus, this choice of operator ordering is not consistent with the existence of virtual black holes. 
 It is interesting to note that we have demonstrated that only 
certain  factor orderings are consistent with the existence of virtual black holes.  

On the other hand at the early time limit  ( $a \rightarrow 2M$ ), 
which means $\alpha \rightarrow 0$ by Eq. (3.26), 
we must consider the cases $p_o<1$ and $p_o>1$ separately. 

When $p_o<1$ the early time limit means $z \rightarrow 0$ , where $z$ is defined in Eq. (3.30). 
Using the relations (4.5) and those which have the argument $-z$ , we obtain from Eq.(3.33) 
$$
\begin{array}{ll}
\dis{I(z)^2 \over D(z)^2} &\sim
\dis{-{\biggl[ {1 \over \sqrt{3}} \vert c_Y \vert^2 
\Bigl( e^{-{\pi \over 3}i}-e^{-{2\pi \over 3}i} \Bigr) 
+c_J c_Y^* e^{-{\pi \over 3}i} -c_J^* c_Y e^{-{2\pi \over 3}i} \biggr]^2 
\over 
\biggl[ {1 \over \sqrt{3}} \vert c_Y \vert^2 
\Bigl( e^{-{\pi \over 3}i}+e^{-{2\pi \over 3}i} \Bigr) 
+c_J c_Y^* e^{-{\pi \over 3}i} +c_J^* c_Y e^{-{2\pi \over 3}i} \biggr]^2}} \\[12mm]
&\sim O(1) \ . 
\end{array}                                                \eqno(4.7)
$$

When $p_o>1$ the early time limit means $z \rightarrow 0$ , where $z$ is defined in Eq. (3.34). 
We use Eq. (3.37) and the relations (4.5) , so we obtain the same result as in the relations 
(4.6), though $z$ is defined in Eq. (3.34).
Hence, both in the cases $p_o<1$ and $p_o>1$ , the black hole would become classical  
at the early time limit, 
since the quantum fluctuations become minimum for  this limit.

\section{Conclusion}

In this paper, we have analysed the effect of quantum fluctuations on the geometry of a black hole. We demonstrated that the quantum fluctuations 
for the black hole depend strongly on the factor ordering chosen. So, for a certain value of the factor ordering parameter 
the quantum fluctuations dominate 
at the Planck scale. However, for anther value of factor ordering parameter, the geometry remains classical even near Planck scale. Hence, only certain 
values of factor ordering parameter are consistent with the proposal of describing spacetime foam in terms of virtual black holes.

It will be interesting to analyse  virtual black holes in some UV complete theory of gravity, like the 
 Horava-Lifshitz     gravity  \cite{3}-\cite{4}. In this theory, space and time have
 different Lifshitz scaling, and it reduces to 
 general relativity in the IR limit. It may be noted that the 
 Wheeler-DeWitt equation for the Horava-Lifshitz    gravity has been studied 
\cite{qabcd1}-\cite{qabcd}, and the third quantization of the 
  Horava-Lifshitz  gravity has also been performed  \cite{mult}. 
  It will be interesting to analyse the Wheeler-DeWitt equation for a black hole in 
Horava-Lifshitz    gravity, and analyse the virtual black hole model using such a 
Wheeler-DeWitt equation. It may also be noted that the modification of
the Wheeler-DeWitt equation from the generalized uncertainty principle 
has also been studied \cite{Majumder}. In this Wheeler-DeWitt equation, 
we can obtain higher derivative corrections. The third quantization 
of such a Wheeler-DeWitt equation for cosmology has already been analysed \cite{faiz}. 
It would be interesting to perform a similar analysis for 
the Wheeler-DeWitt equation of a black hole. 
It has also been argued that  virtual black holes can lead to a vanishing of the QCD $\theta$ parameter \cite{ha}. 
This occurs as virtual black holes can produce  a loss of coherence between the different
vacuum states, and this can lead to the vanishing of the   QCD $\theta$ parameter. It would be interesting to derive this result using the 
third quantized formalism.

 \end{document}